\documentclass{article}
\usepackage{spconf,amsmath,epsfig}
\usepackage{algorithm}
\usepackage{algorithmic}
\usepackage{graphicx, subfigure}
\usepackage[table]{xcolor}

\title{E-HORM: An Energy-efficient Hole Removing Mechanism\\ in Wireless Sensor Networks}

\name{M. B. Rasheed$^{\ddag}$, N. Javaid$^{\ddag}$, Z. A. Khan$^{\$}$, U. Qasim$^{\sharp}$, M. Ishfaq$^{\S}$}

\address{$^{\ddag}$COMSATS Institute of Information Technology, Islamabad, Pakistan. \\
        $^{\$}$Faculty of Engineering, Dalhousie University, Halifax, Canada.\\
        $^{\sharp}$University of Alberta, Alberta, Canada.\\
        $^{\S}$King Abdulaziz University, Rabigh, Saudi Arabia.}
\begin{document}
%\ninept
%
\maketitle
\begin{abstract}
Cluster based routing protocols for Wireless Sensor Networks (WSNs) have been widely used for better performance in terms of energy efficiency.
Efficient use of energy is challenging task of designing these protocols. Energy holes are created due to quickly drain the energy of a few nodes due to non-uniform node distribution in the network. Normally, energy holes make the data routing failure when nodes transmit data back to the sink. %This paper describes the occurrence of energy holes in Low Energy Adaptive Clustering Hierarchy (LEACH), Threshold sensitive Energy Efficient sensor Network protocol (TEEN), Distributed Energy Efficient Clustering (DEEC) and Stable Election Protocol (SEP).
We propose \underline{E}nergy-efficient \underline{HO}le \underline{R}emoving \underline{M}echanism (E-HORM) technique to remove energy holes. In this technique, we use sleep and awake mechanism for sensor nodes to save energy. This approach finds the maximum distance nodes to calculate the maximum energy for data transmission. We consider it as a threshold energy $E_{th}$. Every node first checks its energy level for data transmission. If the energy level of node is less than $E_{th}$, it cannot transmit data. %We also explain mathematically the energy consumption and average energy saving of sensor nodes in each round. Extensive simulations show that when use this approach for WSNs significantly helps to extend the network lifetime and stability period.

%Cluster based routing protocols for Wireless Sensor Networks (WSNs) are widely used for better performance in terms of energy efficiency. Energy efficiency is one of the vital aspect in WSN. Energy holes are created due to non-uniform distribution of nodes throughout the WSN, cause quick drains of nodes energy.  Energy holes make data routing failure from node to sink. To remove energy holes, we proposed \underline{E}nergy-efficient \underline{HO}le \underline{R}emoving \underline{M}echanism (E-HORM) that used sleep awake mechanism to save energy. In this approach, first we find the node that is at maximum distance from the sink. Energy required to transmit data from that node considered to be threshold energy E_{th}. If energy levels of nodes less than E_{th}, it cannot transmit data.
\end{abstract}
\section{Background}
\label{sec:intro}
WSNs consist of a large number of sensor nodes deployed in a sensor field for object monitoring either inside the field or near the field. Recent advances in Microelectromechanical Systems (MEMS) based technologies have enabled the deployment of a large number of tiny sensor nodes with limited battery life time. These nodes have the low computational ability and small internal memory. These small sensor nodes capable of monitoring, sensing, aggregation and transmission of data to the sink. WSNs are used in many communication applications including security, medical, surveillance and weather monitoring.
Sensor nodes are able to measure various parameters of the environment and transmit collected data to the sink directly or through multihop communication.
Nodes deployment is the first step in establishing sensor network. Sensor nodes are battery powered and randomly deployed in target area. After the deployment sensor network cannot perform manually. Optimizing the energy consumption is one of the major tasks in WSNs to prolong the network lifetime. To address this issue, much work has been done in this area during the last few years. If the sensor nodes are deployed uniformly, nodes near the sink send their own data as well as the date collected by other nodes away from the sink in multihop scenario. In this case, the sensor nodes near the sink consume more energy and die out quickly. As a result, the sensor network will disconnect having sufficient energy left unused \cite{1}.
%In this work, we investigate and try to remove the energy holes problem. We analyze the energy imbalance in these protocols and present sleep and awake mechanism to enhance the network lifetime for many to one WSNs.% We use sleep and awake process to eliminate the energy hole problem.
%Further we conduct extensive simulations to investigate and confirm the performance of these techniques. %Simulation results show the network lifetime and the total number of nodes in the network area use to enhance the network lifetime.
%\section{Related Work and Motivation}
%\label{sec:format}
Various schemes have been proposed to address the Energy Hole Problem (EHP). In \cite{1}, authors present a model for balanced density control to avoid energy holes. They use equivalent sensing radius and pixel based transmission schemes for balanced energy consumption. By activating different energy layered nodes in non-uniform distribution the energy holes problem is mitigated effectively.

In \cite{2}, authors purpose a Voronoi diagram-based distribution model for sensor deployment. In this model, each node calculate its Voronoi polygon to detect coverage hole and move towards a better position for maximum coverage in the field.

In \cite{3}, authors discuss the Corona-based sensor network model for balanced energy depletion due to many to one communication in multihop sensor networks. They use mobile sensors to heal the coverage hole created due to large data relaying near the sink. In \cite{4}, authors purpose a multiple sink model to divide the network load near the sink to avoid the energy hole. Decision of a multiple sink is based on the amount of data loads in sensor networks.

In \cite{5}, authors discuss the distributed localization problem, Optimal Geographic Density Control (OGDC) for full coverage as well as connectivity. They prove that if communication range is twice the sensing range then complete coverage implies connectivity.

Tang and Xu \cite{2}, discuss how to optimize the network lifetime and data collection at the same time. Large amount of data is given to the sink by nearby nodes and less data from the nodes that are far away from the sink in the previous study. For this a rate allocation algorithm Lexicographical Max-Min (LMM) for data gathering is proposed to maximize the data-gathering amount and maximize the network lifetime under balance data gathering condition.

\section{Energy hole problem in WSNs}
\label{sec:pagestyle}
%There are many phenomenons that affect the functionality; sensing and communication in WSNs.
In this section, we discuss the characteristics and effects of energy holes. Unbalanced energy consumption is a major issue in WSNs when nodes are randomly deployed in sensor networks. Sensor nodes in the network behave as a data originator and data router \cite{6}.

Nodes near the sink have a greater load of data, hence consume more energy. Therefore, nodes near the sink deplete more energy and die quickly, leading what is called EHP around the sink. In this situation, no more data will be transmitted to the sink. So, the network lifetime ends due to more depletion of energy near the sink. More sensor nodes due to dense deployment in any region may overlap and increase the hardware cost. However dense deployment is another reason for the creation of holes problem in WSNs.\\ %In all current routing schemes using optimal path intermediate nodes in the routing path deplete their energy more quickly, which expand the area of an energy hole.\\
%Using proper nodes deployment techniques, EHP can be reduced and thus extends the network lifetime of WSNs. EHP can be reduced through different node deployment techniques, mobile sink node ~\cite{wu2006energy}, non uniform deployment ~\cite{olariu2006design}, or variable transmission ranges ~\cite{ren2008leds},~\cite{song2008mitigating}, but these deployment strategies bring much more management cost.

%\begin{table}[ht]
%  \centering
%  \caption{\textbf{Cluster Based Protocols}} \vspace{0.5cm}
%  \begin{tabular}{|l|l|} \hline
%    %\toprule
%          %&  Main features of these protocols &&\\
%     \textbf{Protocols} & \textbf{Main features}\\ \hline
%    %\midrule
%    %Cluster based protocols & Main features\\
%    LEACH& Equal cluster radius; each node has the same \\&probability as the cluster head; the cluster head\\&sends data to the sink directly. \\ \hline
%    TEEN & Equal cluster radius; each node has the same \\&probability as the cluster head; the cluster head \\&sends data to the sink directly. \\ \hline
%    DEEC &  Equal cluster radius; multilevel heterogeneous \\&network; advance nodes have more probabilities \\&as the cluster head compare to normal nodes; \\&the cluster head sends data directly to the sink.\\ \hline
%    SEP & Equal cluster radius; two-level heterogeneous \\&network; advance nodes have the more probability \\&as a cluster head as compare to normal nodes; \\&cluster heads send data directly to the sink. \\ \hline
%    %\midrule
% \end{tabular}%
% \label{tab:addlabe1}%
% \end{table}

\section{Energy consumption model}
\label{sec:typestyle}
We have found that due to EHP network dies early. %In ~\cite{li2005analytical}, authors show that due to more depletion of energy near the sink nodes die more quickly, and network lifetime is over, even when up to 90\% of energy is left unused.
\textit{Yifeng et al.} \cite{7} deals with the lifetime of sensor networks. They assume that the communication between nodes consume more energy rather than data aggregation and data reception. Previous research shows that the traffic near the sink is heavier than the traffic away from the sink. Energy consumption near the sink is greater and energy holes occur leading to the death of the entire network. This phenomenon reduces the lifetime of the whole network due to large energy consumption near the sink. If more nodes are deployed near the sink, there will be more nodes use to relay the distant data and hence extends the network lifetime which is also non-uniform node distribution strategy.
The phenomenon of an energy hole makes the researchers realize that the network lifetime is determined by the weakest node.% while the region, size and time of energy holes are the temporal characteristics of the network life time.
How to avoid EHP becomes an important research area now a days.\\
We use the same energy consumption model as used in \cite{8}, which is the first-order radio model. %In this model radio dissipates $E_{elec}= 50nj/bit$ to run the transmitter circuitry, and $E_{amp}=100 pj/bit/m^2$ for the transmitter amplifier. %The radio has power control and can adjust power according to the distance. We use free space $E_{fs}$ and multipath $E_{amp}$ loss model in our scheme.
%Receiving is also a high cost operation therefor; number of transmission and receiving should be minimal.
%To receive the 1-bit message the radio expends $E_{Rx}= LE_{ele}$ energy. LEACH, TEEN, DEEC and SEP use the same constants ($k,E_{fs},E_{amp}$) for calculating energy cost.
\section{E-HORM: Proposed Scheme}
\label{sec:typestyle}
We consider the scenario where nodes are randomly deployed in a given region. Some nodes are selected to be active and rest are in sleep mode to maintain sensing, coverage and connectivity.% We randomly deploy $N$ number of sensors in a square area.
The position of the sink is at the center of the network. Energy of all the nodes are equal while the energy of the sink is unlimited. In our model, nodes transmit data to the sink based on residual energy and the distance between nodes and sink.
%\begin{figure}[ht]
%\begin{center}
%\includegraphics[scale=0.45]{InitialDeployment1.eps}
%\caption{100-node random network}
%\end{center}
%\end{figure}
E-HORM scheme has four major phases: (i) initializing phase, (ii) threshold calculating phase, (iii) cluster formation, and (iv) sleep/awake scheduling phase. In every round sink first checks the maximum distance node in the field. It then calculates the required energy to transmit data to the sink. We set this energy as a $E_{th}$. In every round if the energy level of a node is greater than or equal to $E_{th}$, sensor node transmits data to the sink. If the energy level of any node is less than $E_{th}$, it cannot transmit data to the sink. When the energy level is less than $E_{th}$ value, it moves towards the sleep mode to save energy. %In the next round again sink calculates the energy level of each node. The same process repeats until the number of sleep nodes are equal to $n_s=10$. Now we discuss the idea of awake mechanism. When numbers of sleep nodes are greater $n_s>10$ than the node which is fist one in the sleep position move towards awake position. In next round when the number of node in sleep position is greater $n_s>11$ than the node which is in second position of the sleep position moves towards the active position and so on.
%When the number of sleep nodes $n_s<10$ than the node whose energy level is less then threshold energy level moves towards sleep position. In this scenario, the total node in the sleep position is again equal to $n_s=10$. Flowchart describes the sleep and awake mechanism.
%\begin{figure}[]
%\begin{center}
%\includegraphics[scale=0.45]{Flowchart.eps}
%\caption{E-HORM Flowchart}
%\end{center}
%\end{figure}
%We implement our scheme in the cluster based protocols LEACH, TEEN, DEEC and SEP. Nodes choose their cluster head on the basis of predefined probability. The cluster heads broadcast their status that each node can determine to which cluster head it wants to associates to consume minimum energy for data transmission. After association, each cluster head creates a schedule to the nodes in its cluster. Sink assigns TDMA slots to every node. Nodes only transmit their data during their assigned TDMA slots. This is to save energy for the nodes that are in sleep mode except during transmission.
\subsection{Sensor Node Sleep Scheduling}
Before performing the sleep schedule, we examine the energy level of each node according to their distance from sink by using the following steps.
\begin{itemize}
  \item Case 1: $E_r>E_{th}$: When $E_r$ is greater than the $E_{th}$, the node is in active mode and ready for communication.
  \item Case 2: $E_r<E_{th}$: When $E_r$ is less than the $E_{th}$, the node moves towards sleep position.
\end{itemize}
%First we calculate the value of threshold energy of each node for sleep and awake mechanism. Sink selects the node to put into sleep position randomly.
Each node set the sleeping scheduling according to the $E_{th}$.\\
To calculate the $E_{th}$, we use the following formula.
\begin{equation}
E_{th}=((ETX+EDA)*D)+(E_{amp}*D*d^4)
\end{equation}
Where $D$ is the length of data packet and $d$ is the distance between maximum distance node and sink.
%Sink calculates the threshold energy for sleep and awake mechanism.% CH2 is the more distant node and consumes more energy for transmission. We calculate the threshold energy of this node according to the given formula.
\subsection{E-HORM Formulation}
Based on the network model, nodes belonging to CH forward both the data generated by themselves and the data generated by its member nodes. Nodes which are not CH need not forward any data. Suppose nodes are randomly deployed in the network and there is no need for data aggregation at any forwarding node. Based on transmission mechanism data for CH receive and forward is $(D_1+D_2+D_3+....+D_N)$ and $(D_{CH}+D_1+D_2+D_3+....+D_N)$.
%\begin{figure}[ht]
%\begin{center}
%\includegraphics[scale=0.45]{Threshold.eps}
%\caption{Threshold energy of distant node}
%\end{center}
%\end{figure}
If the distance between N and CH is $d<d_0$ than energy consumption for data transmission from N to the CH.
\begin{equation}
E_N^{CH}= D_N^{CH}(E_{ele})+D_N^{CH}(E_{fs})(d^2)
\end{equation}
%where $d$ is the distance between N and CH.
%The nodes send their own data to CH for each data-gathering process in different rounds.\\
Now considering the scenario where distance between N to CH is $d>d_0$.
\begin{equation}
E_N^{CH}= D_N^{CH}(E_{ele})+D_N^{CH}(E_{amp})(d^4)
\end{equation}
Energy consumed by CH to transmit data to the S when distance between them is $d<d_0$:
\begin{equation}
  E_{CH}^S= D_{CH}^S(E_{ele})+E_{DA}+D_{CH}^S(E_{fs})(d^2)
\end{equation}
When distance between CH and S is $d>d_0$:
\begin{equation}
  E_{CH}^S= D_{CH}^S(E_{ele})+E_{DA}D_CH^S(E_{amp})(d^4)
\end{equation}
\begin{equation}
  E_{{Total}_-{CH}}=E_{CH}+E_N
\end{equation}
\begin{equation}
  E_{{Average}_-{CH}}=\frac{E_{{Total}_-{CH}}}{N}
\end{equation}
Energy saving in each round for normal node is:
\begin{equation}
  E_{{Save}_-{N}}=E_{elec}+E_{TX}+E_{amp}
\end{equation}
Energy saving for CH is:
\begin{equation}
  E_{{Save}_-{CH}}=E_{ele}+E_{DA}+E_{TX}+E_{RX}+E_{amp}
\end{equation}
Energy saving for all sleep nodes
\begin{equation}
  E_{{Save}_-{Total}}=\sum\limits_{i=0}^n E_i
\end{equation}
%Where $n$ are the total number of nodes that are in sleep mode\\
%Average energy saving is
%\begin{equation}
%  E_{{Save}_-{Average}}=\frac{E_{{Save}_-{Total}}}{N}
%\end{equation}
%In each round CH receive data and aggregate before transition to the sink. Additional energy is required for data aggregation, so CH consumes more energy as compared to the normal nodes.
%Fig. 3 shows CHs with different distances from the sink. Each cluster head contains different numbers of sensor nodes. CH1 shows two regions. Distance between CH and N is less then $d_0$ in the dotted region. Energy required to transmit data from N to CH is less than the energy required for the node that is outside from the dotted region. This is because the distance between N outside the dotted region is greater than $d_0$. Energy required to transmit data from CH1 to S is greater because the distance is $d>d_0$. In CH7 distance between the S and CH is less than $d<d_0$. Energy used to transmit data from CH7 to S is less than CH1.
\subsection{Analysis}
Sleep probability of sensor nodes for each round is based on $E_{th}$ of distant nodes. The nodes away from the sink increase the sleep probability. In this way, the nodes consume approximately balance energy to enhance the network lifetime.
For WSNs the sleeping scheduling is very important due to limited energy of sensor nodes. If a node set into active position for a long time, it consumes a lot of energy. On the other way, the transmissions create more delay for long time sleep duration. In this paper, we design an optimum sleeping control mechanism to avoid both of the problems.

\begin{figure}
\centering
\mbox{\subfigure{\includegraphics[width=1.7in]{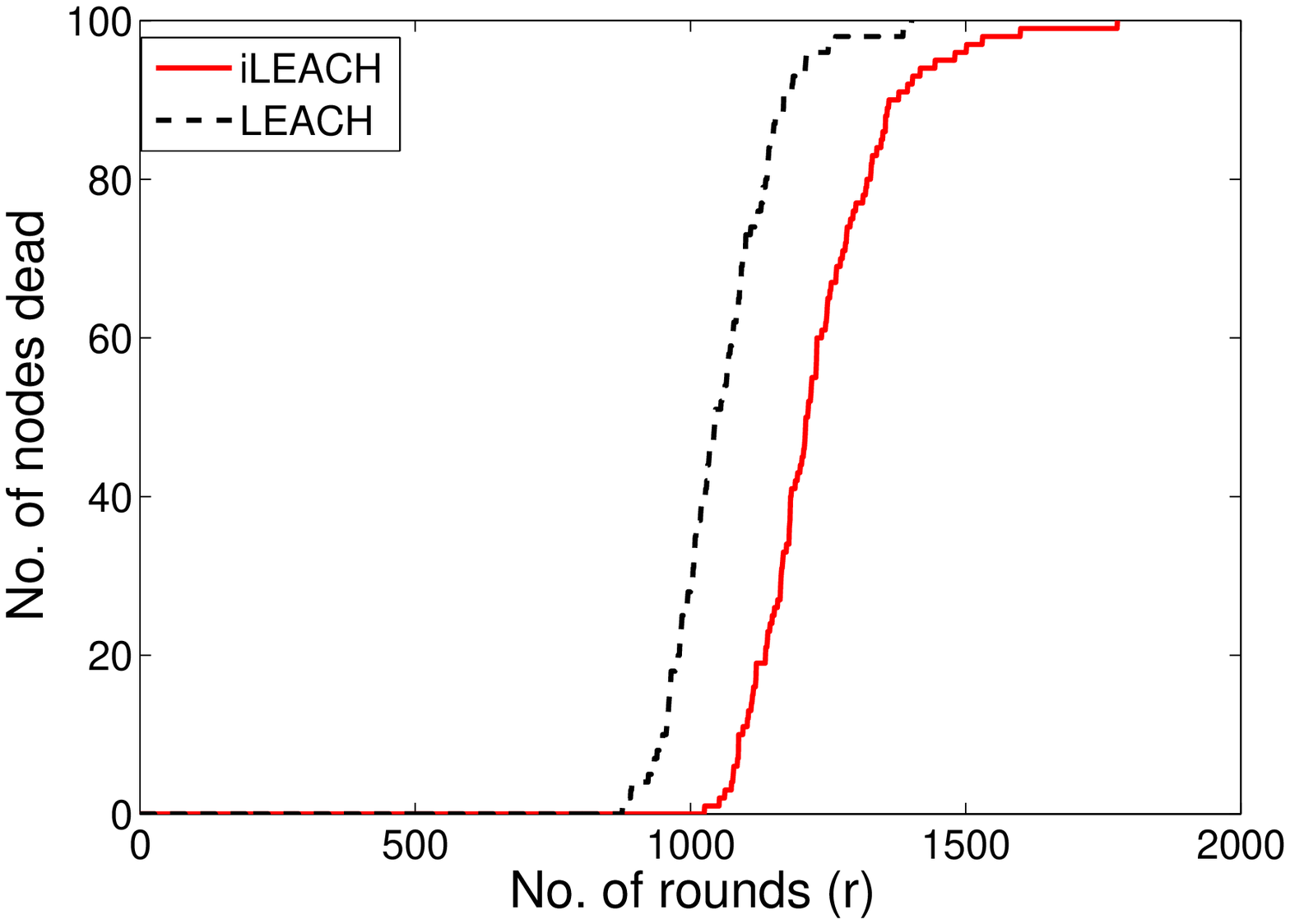}}\quad
\subfigure{\includegraphics[width=1.7in]{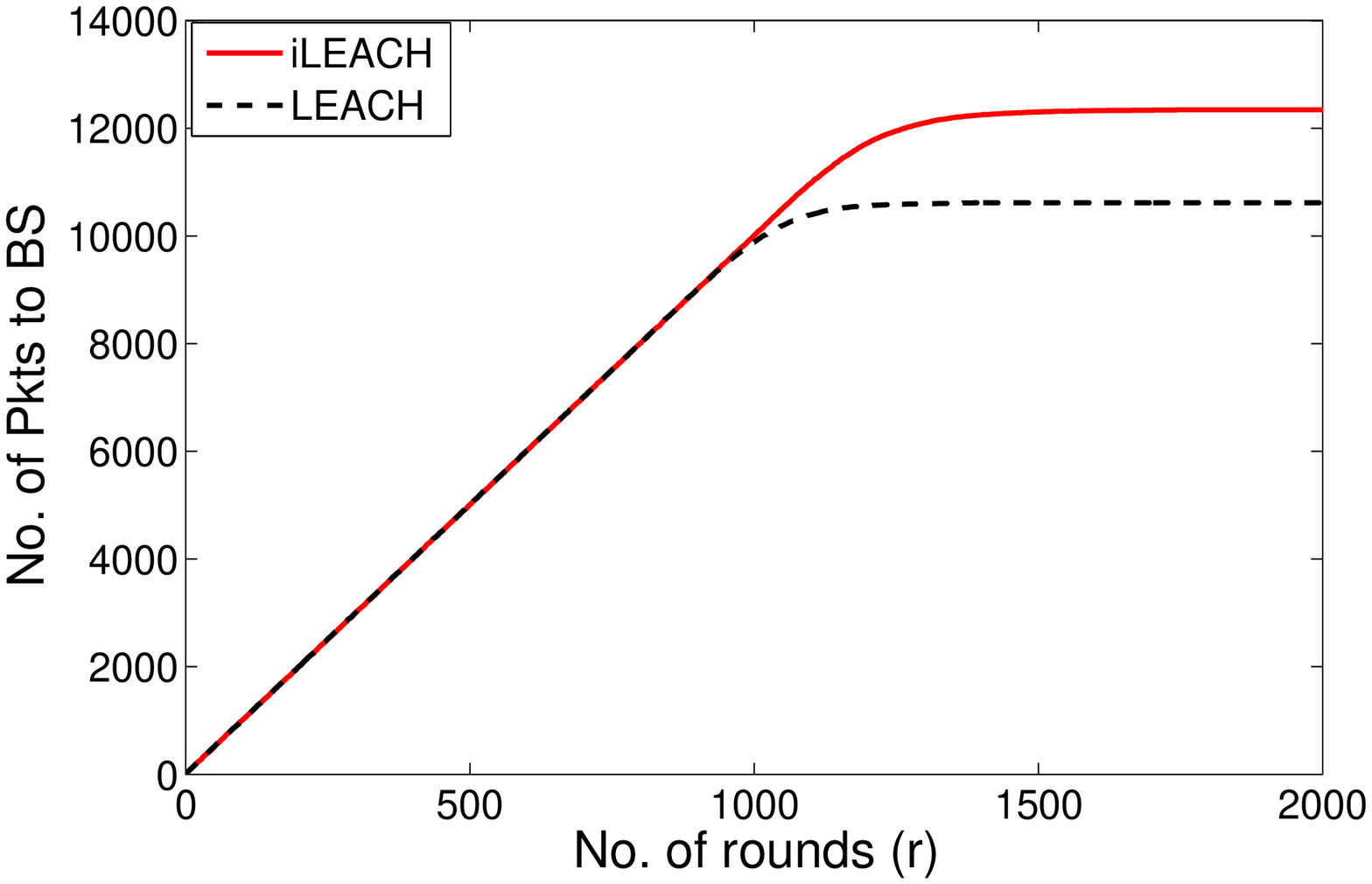} }}
\caption{Comparing the performance of LEACH and iLEACH} \label{fig12}
\end{figure}

\begin{figure}
\centering
\mbox{\subfigure{\includegraphics[width=1.7in]{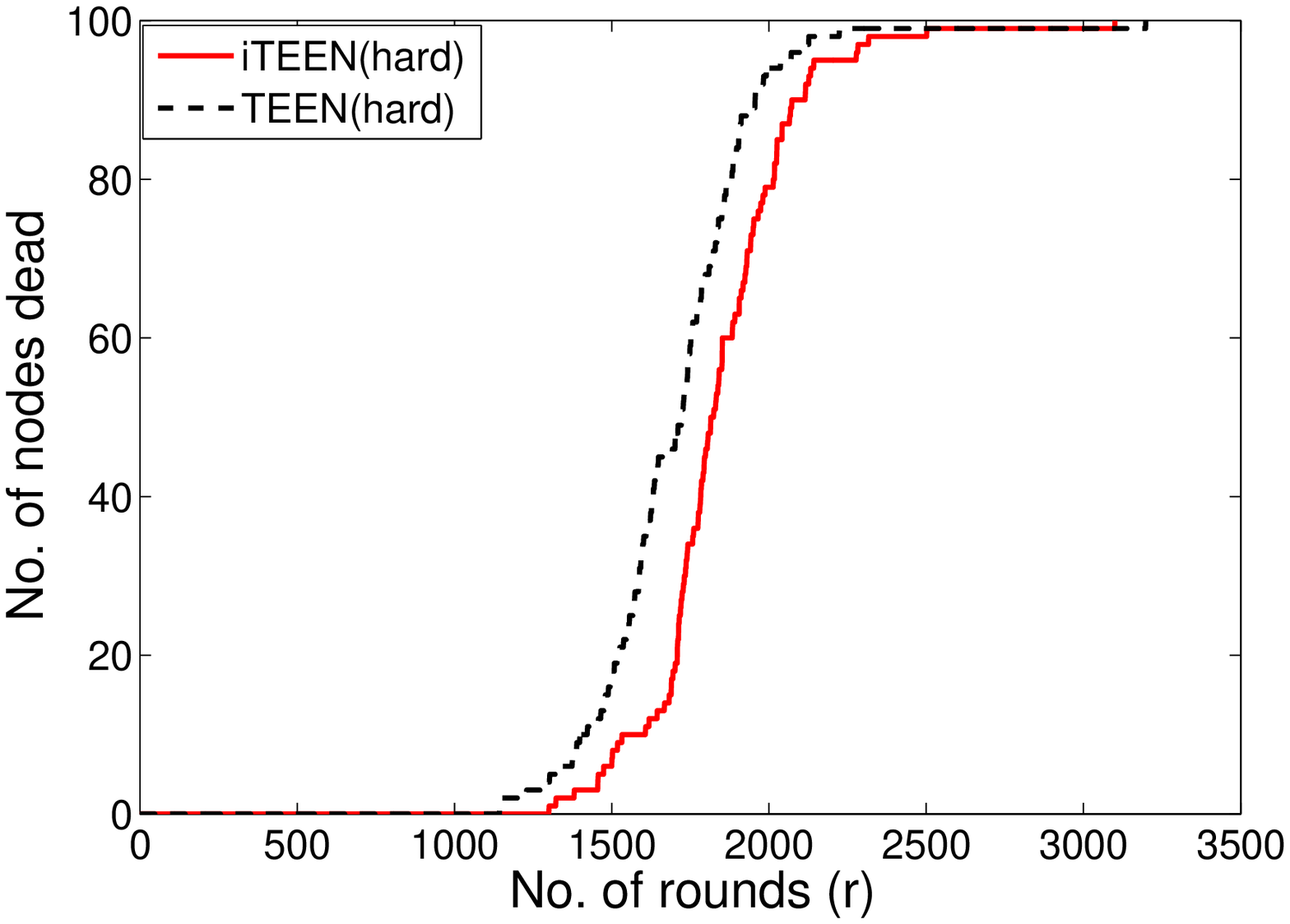}}\quad
\subfigure{\includegraphics[width=1.7in]{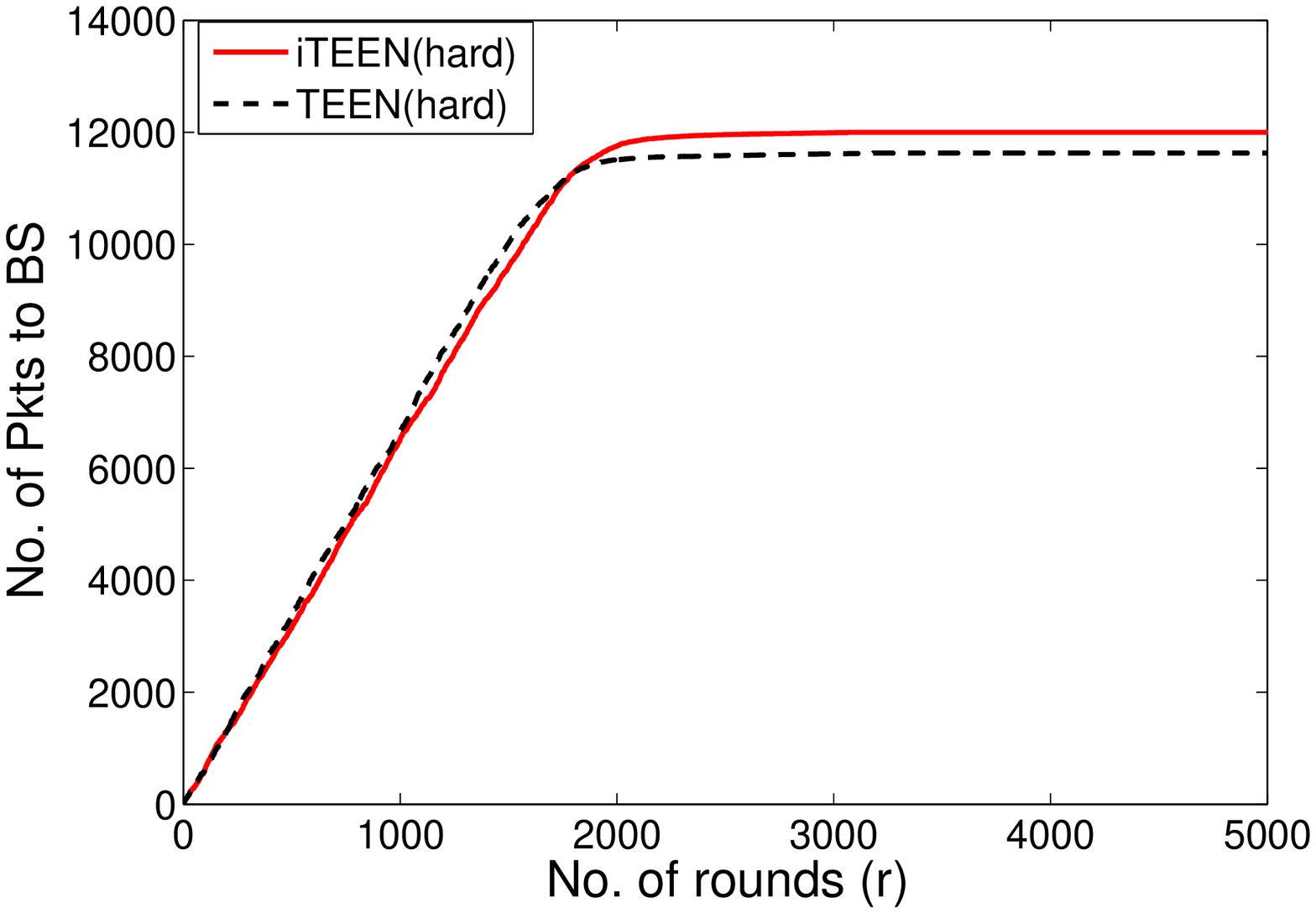} }}
\caption{Comparing the performance of TEEN and iTEEN} \label{fig12}
\end{figure}

\section{Simulation results}
%\subsection{Network Model}
Our results are based on analyses and are validated by Matlab simulations. In our simulations, we consider a sensor network with $n$ number of homogeneous or heterogeneous sensor nodes which are randomly deployed in a square field. The only sink is located at the center of the field. Sensor nodes do not move after deployment. Sensor nodes are limited initial energy $E_{init}$, while the energy of the sink is unlimited. The transmission ranges of sensor nodes are adjustable according to the distance from the sink. All nodes need to send the data packets to the sink in a cycle time. In each round sensor nodes are selected to work and the rests of nodes are set to sleep mode to save energy. In E-HORM, this mechanism is called sleep awake process. We apply our scheme in four categories of the cluster based protocols LEACH, TEEN, DEEC and SEP.\\ %In this paper the cluster based network refers to the routing protocols.\\
%All parameter values use in our simulations are listed below.
\begin{table}[h!]
  \centering
  \caption{\textbf{Simulation Parameters}}  \vspace{0.5cm}
  \tiny
  \begin{tabular}{|l|l|l|}\hline
  \textbf{Symbol} &\textbf{Description}   &\textbf{Value}\\ \hline
    $X_m$  & Distance at x-axes &100 meter \\ \hline
    $Y_m$  &Distance at y-axes  &100 meter\\ \hline
     N     &Total number of nodes & 100 Nodes \\ \hline
   $E_0$   &Total energy of node & 0.5 j \\ \hline
    $P$    &Probability of cluster head & 0.1 \\ \hline
    $E_{RX}$ &Energy dissipation: receiving & $0.0013/pj/bit/m^4$\\ \hline
    $E_{fs}$ &Energy dissipation: free space model & $10/pj/bit/m^2$\\ \hline
    $E_{amp}$&Energy dissipation: power amplifier & $100/pj/bit/m^2$ \\ \hline
    $E_{ele}$ &Energy dissipation: electronics   & $50nj/bit$ \\ \hline
    $E_{TX}$  &Energy dissipation: transmission   & 50/nj/bit \\ \hline
    $E_{DA}$  &Energy dissipation: aggregation    & $5/nj/bit$ \\ \hline
     $d_0$    &Reference distance                 & 87 meter  \\ \hline
  $n$        &Number of sleep nodes               & 10 Nodes\\ \hline
    $E_r$        &Remaining energy of nodes               & \\ \hline
    \end{tabular}%
  \label{tab:addlabel}%
\end{table}%

%Distant cluster heads consumes more energy during transmission.
\begin{figure}
\centering
\mbox{\subfigure{\includegraphics[width=1.7in]{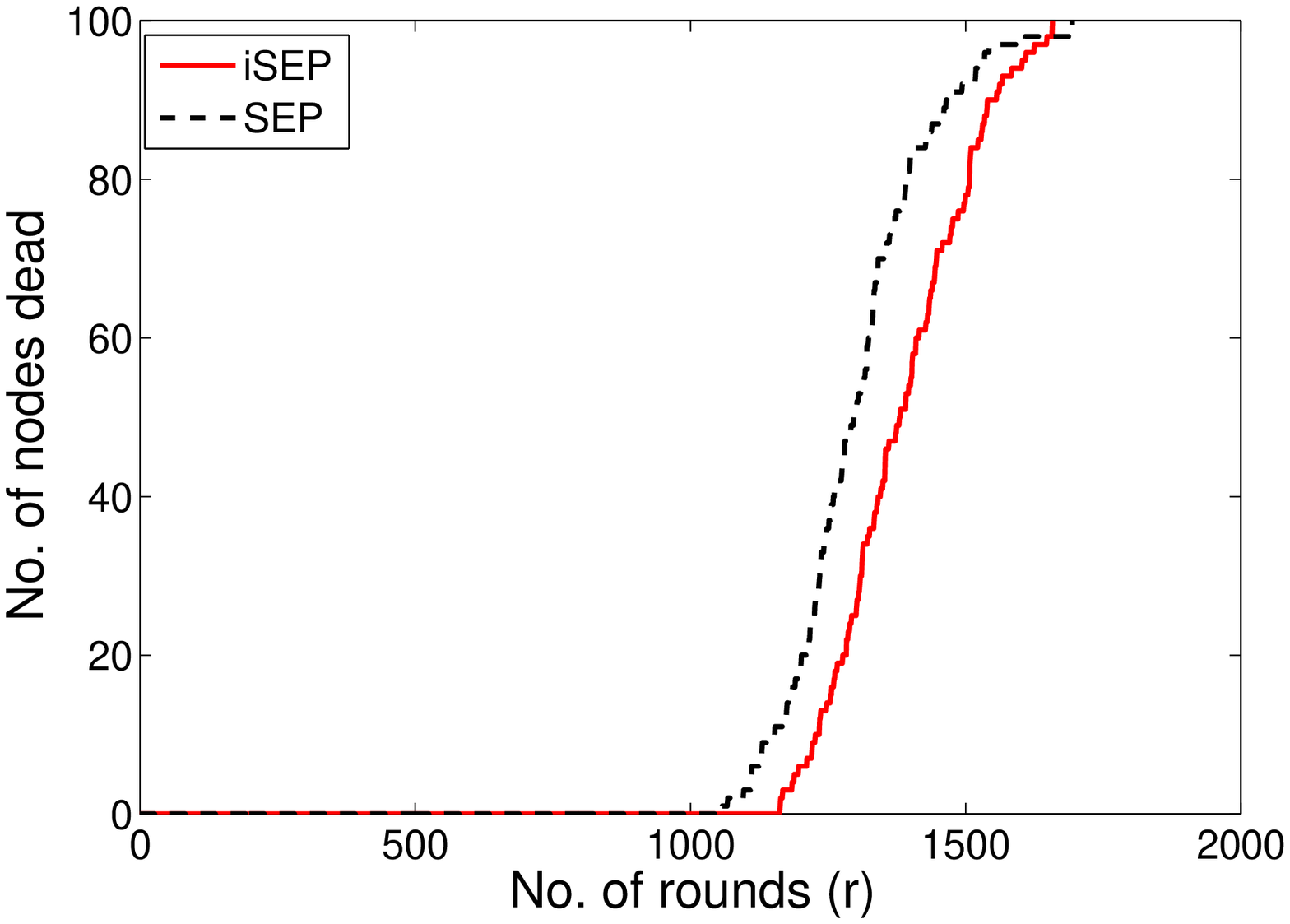}}\quad
\subfigure{\includegraphics[width=1.7in]{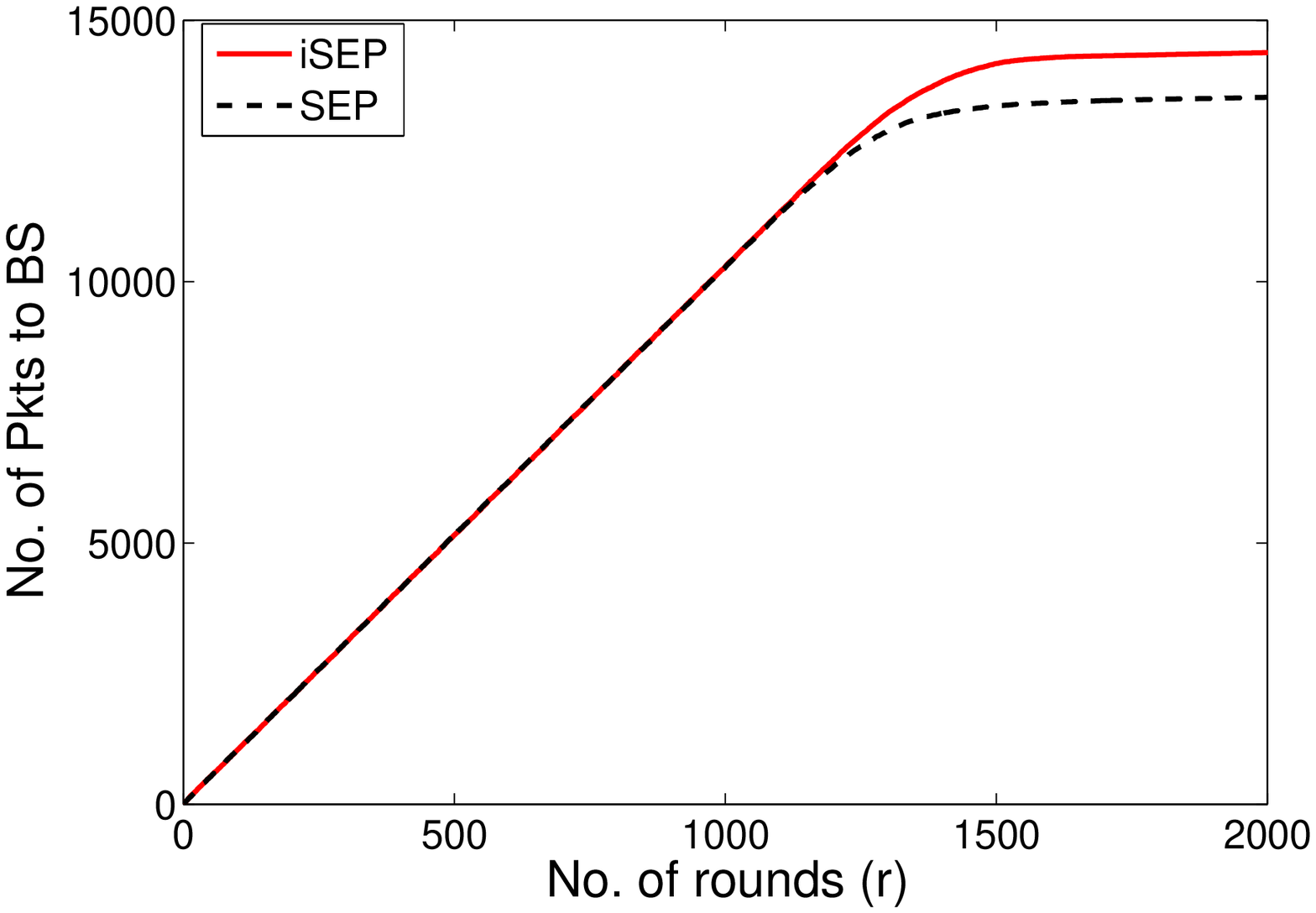} }}
\caption{Comparing the performance of SEP and iSEP} \label{fig12}
\end{figure}
%\vspace{-4.5mm}
\begin{figure}
\centering
\mbox{\subfigure{\includegraphics[width=1.7in]{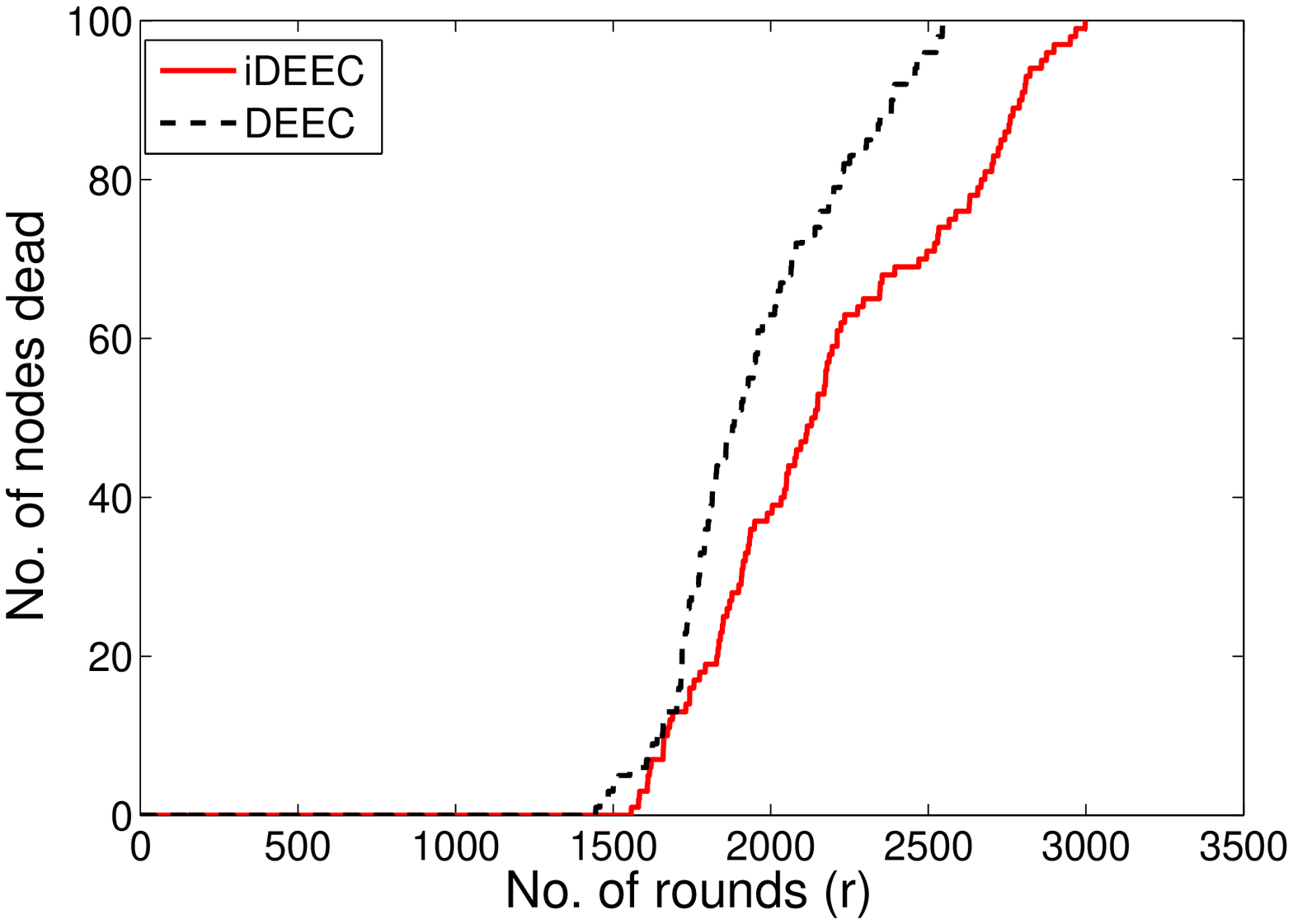}}\quad
\subfigure{\includegraphics[width=1.7in]{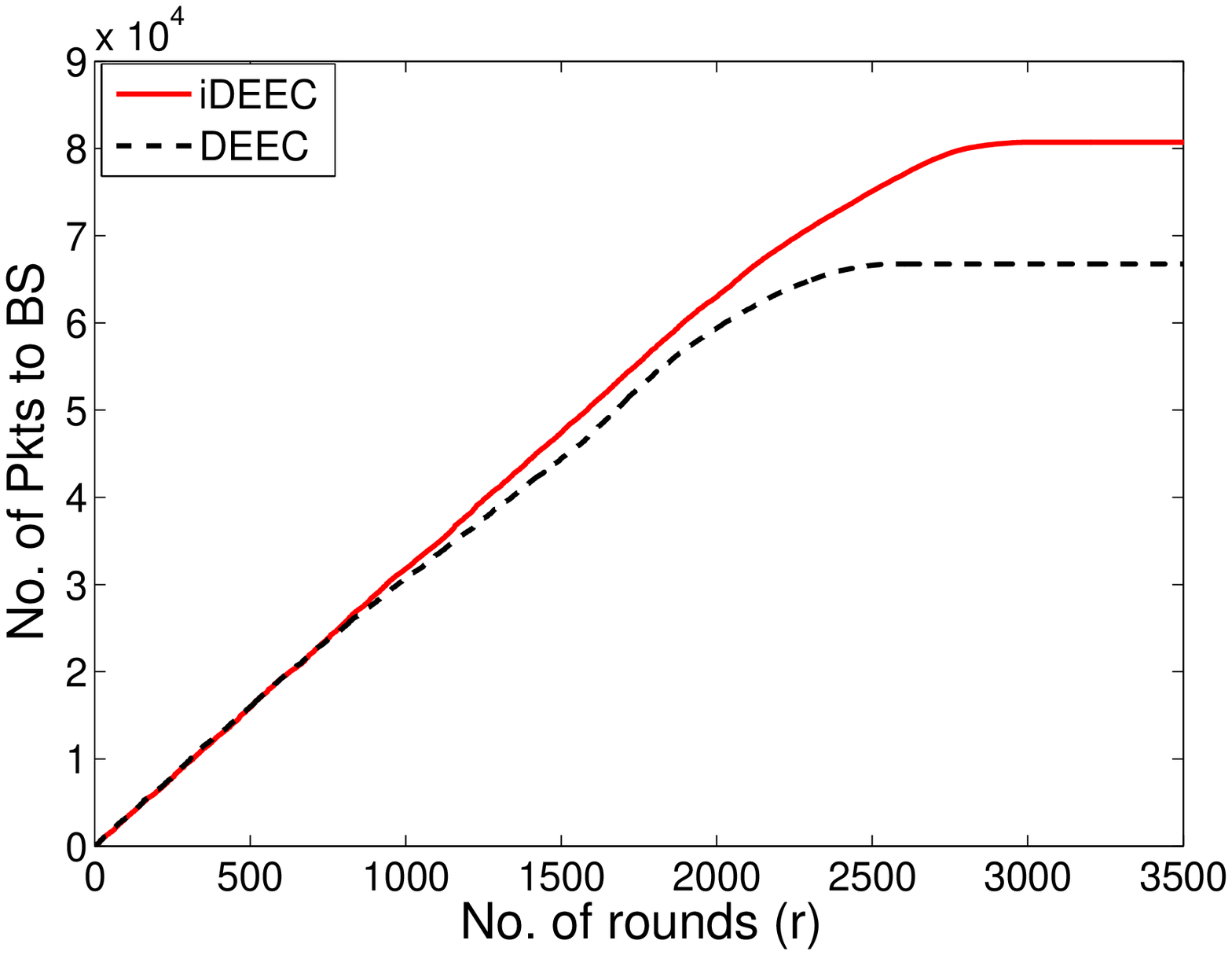} }}
\caption{Comparing the performance of DEEC and iDEEC} \label{fig12}
\end{figure}
In this section, we evaluate and explain the simulation results of our purposed work. Three main matrices, including the stability period, network survival lifetime and data transmitted to the sink are calculated and compared with the existing work. %We write Matlab program to simulate our work with initial deployment of nodes.
We explore the theoretical analysis of node distribution according to random deployment. We evaluate and compare our purposed technique with LEACH, TEEN, DEEC and SEP protocol with random node deployment in the sensor network. Our technique outperforms in terms of network life time and stability period. Now we explain how the sleep and awake mechanism are carried out in LEACH and TEEN to remove the energy holes. LEACH and TEEN are homogeneous routing protocol and all the nodes have the same probability to become a cluster head. \\%Cluster head consumes more energy then normal nodes.\\
Cluster head consumes more energy during transmission due to data load of member nodes. Nodes forward their own data to cluster heads according to TDMA schedule. All nodes are in sleep node and turn on their transmitters during data transmission to save energy. According to our approach, the nodes that have the energy level less than the threshold are in sleep mode to save energy. In this way we save energy to prolong network life time and stability period.% Simulation results of iLEACH and LEACH is shown in above figures.
%The network lifetime, stability period and data transmitted to the sink is greater in iLEACH.
%All nodes participate in the communication process.\\
%We find that it is difficult to remove all energy holes due to random deployment nature of sensor nodes in WSNs. We describe that with the purposed transmission scheme, better results are achieved.
%The initial energy for each node is $\varepsilon$ is $0.5 j$. The length of data is $L$ is $4000$ bits. Fig. 2 shows the initial node deployment of sensor nodes.

%\begin{figure}[ht]
%\begin{center}
%\includegraphics[scale=0.3]{iVsDeecAlive.eps}
%\caption{Number of alive nodes in different target areas}
%\end{center}
%\end{figure}
%\begin{figure}[ht]
%\begin{center}
%\includegraphics[scale=0.3]{iVsDeecPkts.eps}
%\caption{Number of alive nodes in different target areas}
%\end{center}
%\end{figure}
SEP and DEEC are the heterogeneous clustering protocol. In this section, we compare the performance of SEP and DEEC with our proposed scheme iSEP and iDEEC in the same heterogeneous setting. Extra energy is distributed over-all advance nodes in the field. This setting latter provide the more stable region and network lifetime. Fig. 3 shows the result of SEP and our proposed scheme iSEP. It is very clear that the stable region of iSEP is greater than SEP even though the network lifetime is not very large. Due to balanced energy consumption network, stability increased. In SEP, when first node dies the system becomes unstable due to population reduction. The death ratio of normal nodes are greater than advance nodes due to having more energy of advance nodes. Distant nodes consume more energy and are put into sleep mode to avoid an energy holes. In each round, the nodes having less energy then $E_{th}$ put into sleep mode to save energy. This mechanism enhances the stability period of SEP due to better utilization of energy.
Now we evaluate and compare the performance, of DEEC and iDEEC protocols. For best evaluation of performance, we ignore the disturbances due to signal collision and interference in a wireless medium. It is obvious that the network lifetime of our proposed scheme is grater compared with DEEC protocols. There is a minute change in the stability period of DEEC and iDEEC while, the network lifetime of iDEEC is greater. This is because the energy of each node is different from other nodes. In periodically sleep and awake mechanism, we can best utilize the energy consumption in each region to remove energy holes.% Due to multilevel heterogeneity sleep awake mechanism
%\begin{figure}
%\centering
%\begin{subfigure}{.5}
%  \centering
%  \includegraphics[width=.4\linewidth]{iVsDeecPkts.eps}
%  \caption{A subfigure}
%  \label{fig:sub1}
%\end{subfigure}%
%\begin{subfigure}{.5}
%  \centering
%  \includegraphics[width=.4\linewidth]{iVsDeecPkts.eps}
%  \caption{A subfigure}
%  \label{fig:sub2}
%\end{subfigure}
%\caption{A figure with two subfigures}
%\label{fig:test}
%\end{figure}

%\begin{figure}
%\centering
%\mbox{\subfigure{\includegraphics[width=1.7in]{iVsDeecPkts.eps}}\quad
%\subfigure{\includegraphics[width=1.7in]{iVsDeecPkts.eps} }}
%\caption{Text pertaining to both graphs ...} \label{fig12}
%\end{figure}
\section{Conclusion}
\label{sec:majhead}
In this article, we focus on energy hole problem and energy consumption in LEACH, TEEN, DEEC and SEP protocols. We discussed the creation of energy holes in homogeneous and heterogeneous routing protocols. We implement our approach in LEACH, TEEN, DEEC and SEP routing protocol. Due to random deployment in these protocols, there exists the probability of energy holes. Sleep and awake mechanism to remove energy holes in WSNs is proposed. We investigated that after our proposed scheme, a better energy consumption is achieved. As for network life time, this work clearly gives the results in terms of network lifetime and stability period. Sensor nodes consume balance energy, and hence maximize the network lifetime. This paper clearly points out how we can remove the energy holes problem in WSNs, and other researchers can also easily purpose a new protocol according to deployment techniques, which avoid the energy holes. Simulation results show that the results of our scheme perform better than previous schemes in terms of network life time and stability perid.


\begin{thebibliography}{00}

\bibitem{1} J. Jia, J. Chen, X. Wang, and L. Zhao, “Energy-balanced
density control to avoid energy hole for wireless sensor
networks,” International Journal of Distributed Sensor
Networks, vol. 2012, 2012.
\bibitem{2} X. Tang and J. Xu, “Optimizing lifetime for continuous
data aggregation with precision guarantees in wireless
sensor networks,” IEEE/ACM Transactions on Networking
(TON), vol. 16, no. 4, pp. 904–917, 2008.
\bibitem{3} J. Jia, X. Wu, J. Chen, and X. Wang, “Exploiting sensor
redistribution for eliminating the energy hole problem in
mobile sensor networks,” EURASIP Journal on Wireless
Communications and Networking, vol. 2012, no. 1, p. 68,
2012.
\bibitem{4} M. Ahadi and A. Bidgoli, “A multiple-sink model for decreasing
the energy hole problem in large-scale wireless
sensor networks,”
\bibitem{5} H. Zhang and J. Hou, “Maintaining sensing coverage and
connectivity in large sensor networks,” Ad Hoc \& Sensor
Wireless Networks, vol. 1, no. 1-2, pp. 89–124, 2005.
\bibitem{6} I. Akyildiz, W. Su, Y. Sankarasubramaniam, and
E. Cayirci, “A survey on sensor networks,” Communications
Magazine, IEEE, vol. 40, no. 8, pp. 102–114, 2002.
\bibitem{7} G. Chen, C. Li, M. Ye, and J. Wu, “An unequal clusterbased
routing protocol in wireless sensor networks,”
Wireless Networks, vol. 15, no. 2, pp. 193–207, 2009.
\bibitem{8} J. Li and P. Mohapatra, “An analytical model for the energy
hole problem in many-to-one sensor networks,” in
IEEE Vehicular Technology Conference, vol. 62, p. 2721,
IEEE; 1999, 2005.
\end{thebibliography}
\end{document}